\documentclass[pra, aps]{revtex4}  

\newcommand{\Tr}{\mathop{\rm Tr}\nolimits} 
\newcommand{\ch}{\mathop{\rm ch}\nolimits} 
 
\usepackage{amsmath,amsfonts,amssymb}
\usepackage{graphicx}
\usepackage{bpchem}

\begin{document} 

\title{Experimental and theoretical investigations of quantum state transfer and decoherence processes in quasi-one-dimensional systems in multiple-quantum NMR experiments}

\author{G. A. Bochkin$^1$, S. I. Doronin$^1$, S. G. Vasil'ev$^1$, A. V. Fedorova$^1$, E. B. Fel'dman$^1$}
\affiliation{$^1$ Institute of Problems of Chemical Physics, Chernogolovka, Moscow Region,
142432, Russia}

\pagestyle{empty} 

\begin{abstract}
Multiple  quantum (MQ) NMR methods \cite{Baum} are applied to the analysis of various problems of quantum information processing.
 It is shown that the two-spin/two-quantum Hamiltonian \cite{Baum} describing MQ NMR dynamics is related to the flip-flop Hamiltonian of a one-dimensional spin system in the approximation of the nearest neighbor interactions.
  As a result, it is possible to organize quantum state transfer  along a linear chain.
	 MQ NMR experiments are performed on quasi-one-dimensional chains of  \BPChem{\^{19}F} nuclei in calcium fluorapatite \BPChem{Ca\_5(PO\_4)\_3F}. Relaxation of the MQ NMR coherences is considered as the simplest model of decoherence processes. A theory of the dipolar relaxation of the MQ NMR coherences in one-dimensional systems is developed. A good agreement of the theoretical predictions and the experimental data is obtained.
\end{abstract}


\maketitle

\section{INTRODUCTION}
\label{sect:intro}  

The development of experimental and theoretical methods of quantum information processing (QIP) is an important direction of quantum informatics. 
 NMR methods are the  simplest ones among different experimental approaches \cite{Nielsen}. 
It is well known that the liquid state NMR methods \cite{Jones} are widely used for experimental realizations of quantum gates and algorithms on the basis of the pseudo-pure states \cite{Cory, Ger}. However, the possibilities of QIP based on the liquid state NMR are limited because the number of correlated qubits is very small (about 10) \cite{Warren} and quantum correlations in such systems are not strong. In particular, entanglement is almost absent \cite{Braun}.

At the same time, the potential of the solid state NMR methods \cite{Feld1} is not exhausted yet.
 Multiple  quantum (MQ) spectroscopy in solids \cite{Baum} is an example of such a method.
 MQ NMR not only creates multi-qubit coherent states but also allows the investigation of  their relaxation under the action of the correlated spin reservoir. MQ NMR \cite{Baum} is an important method for the  investigation of various problems of quantum information processing such as the transmission of quantum information \cite{Cappellaro} and decoherence processes \cite{Doronin,Kaur}.

One-dimensional MQ NMR methods are very suitable for solving problems of quantum informatics.
 The reason is that a consistent quantum-mechanical theory of MQ NMR dynamics has been developed only for one-dimensional systems \cite{Feld2,Feld3,Doronin2,Feld4}. That theory is based on the fact that the non-secular two-spin/two-quantum Hamiltonian \cite{Baum}, describing MQ NMR dynamics, is the XY Hamiltonian \cite{Mattis}, which can be diagonalized exactly for one-dimensional systems in the approximation of the nearest neighbor interactions \cite{Abragam}. As a result, MQ NMR dynamics in such systems can be studied analytically. In particular, only MQ NMR coherences of the zeroth and plus/minus second orders arise in a one-dimensional chain initially prepared in a thermodynamic equilibrium state \cite{Feld2, Feld3,Doronin2,Feld4}.

The developed theory \cite{Feld2, Feld3,Doronin2,Feld4} is based on the model of an isolated spin chain with the nearest neighbor interactions. 
Such a model is realized in quasi-one-dimensional chains of  \BPChem{\^{19}F} nuclei in calcium fluorapatite \BPChem{Ca\_5(PO\_4)\_3F} with a hexagonal system of fluorine nuclei.
The distance between neighboring chains is about three times larger than the distance between  the nearest fluorine nuclei in the same chain.
Taking into account that the dipole-dipole interaction (DDI) of the next nearest spins in the chain is eight times weaker than the nearest neighbor interaction \cite{Abragam} one can use the model of an isolated spin chain with the nearest neighbor interactions for the interpretation of the experimental data obtained from the MQ NMR experiments \cite{Baum}.

The information resource of MQ NMR exceeds the resource of conventional NMR \cite{Feld4}.
However, that resource can be used only  if the  XY Hamiltonian is created with sufficient  accuracy \cite{Doronin3}.
The XY Hamiltonian is obtained as the  averaged two-spin/two-quantum Hamiltonian \cite{Baum} which is the average Hamiltonian \cite{Haeberlen} in the MQ NMR  experiments. 

The corrections to the average Hamiltonian \cite{Haeberlen} spoil the XY Hamiltonian. These corrections  are proportional to powers of the parameter  $\epsilon=t_c \omega_{loc}$ ($t_c$ is the period of the irradiating sequence of the preparation period of the MQ NMR experiment and $\omega_{loc}$ is of the order of the DDI). We have designed a special probe \cite{Doronin3} for  the generation of high-power ultra-short pulses and  decreased  the period $t_c$ and  the parameter  $\epsilon$ in MQ NMR experiments. The results \cite{Doronin3} show  high accuracy of creating the XY Hamiltonian.

It is very important that the XY Hamiltonian (the MQ NMR Hamiltonian) is related to the flip-flop one by a simple unitary transformation \cite{Doronin2} for one-dimensional systems in the approximation of the nearest neighbor interactions. That fact allows us to realize the quantum state transfer along a one-dimensional spin chain. 
It is also significant that the relaxation of the MQ NMR coherences can be considered as a model for studying  the decoherence processes.

The paper is organized as follows. 
An introduction to MQ NMR dynamics is given in section~2. In section 3 we show that MQ NMR  can be used for the quantum state transfer in one-dimensional spin chains.
Relaxation of the MQ NMR coherences in one-dimensional spin chains is discussed in section 4. 
We briefly summarize our results in section 5.

\section{MQ NMR dynamics of one-dimensional spin systems}

\begin{figure}
\begin{center}
\begin{tabular}{c}
\includegraphics[height=5.5cm]{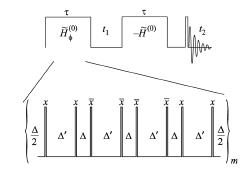}
\end{tabular}
\end{center}
\caption 
{ \label{fig:MQNMR}
Scheme of the multiple quantum NMR experiment. The nonsecular two-spin/two-quantum Hamiltonians $H^{(0)}$, $H^{(0)}_\Phi$  are determined by Eqs. (\ref{averaged},    \ref{twotwo}); $\tau$ is the duration of the preparation and mixing periods; $t_1$ and $t_2$ are durations of the evolution and detecting periods, respectively.
The basic cycle  of the multi-pulse sequence consisting of eight $\pi/2$ pulses with the duration $t_p$. The pulses are separated by delays $\Delta$ and $\Delta'=2\Delta+t_p$, repeated $m$ times for obtaining the necessary time $\tau=12m$ ($\Delta + t_p$); $m$ is a positive integer.}
\end{figure} 

The MQ NMR experiment \cite{Baum} consists of four distinct periods of time (Fig. \ref{fig:MQNMR}): preparation($\tau$), evolution ($t$), mixing($\tau$) and detection.
The spin system is irradiated by a periodic sequence of resonance radio-frequency pulses on the preparation period.
As a result, the anisotropic  DDI \cite{Goldman} become rapidly oscillating if the inverse period of the irradiating sequence $t_c^{-1}$ strongly exceeds the magnitude of the interactions  $\omega_{loc}$ ($\epsilon=t_c \omega_{loc}\ll 1$). 
The theory of the average Hamiltonian \cite{Haeberlen} makes it possible to find the averaged interactions, the Hamiltonian of which $H^{(0)}$ (the averaged non-secular two-spin/two-quantum Hamiltonian \cite{Baum}) up to the terms of the order of $\epsilon^2$ has the form
   \begin{equation}
   \label{averaged}
   H^{(0)}=H^{(+2)}+H^{(-2)},
   \end{equation}
where the non-secular DDI $H^{\pm 2}$ are as follows:
   \begin{equation}
     H^{(+2)}=-\frac{1}{2}\sum\limits_{i,j}D_{ij} I_i^+ I_j^+, \qquad H^{(-2)}=-\frac{1}{2}\sum\limits_{i,j}D_{ij} I_i^-I_j^-.
   \end{equation}
Here $I^\pm_i=I_{ix}\pm I_{iy}$, $I_{i\alpha}$ is the projection of the spin angular momentum of spin $i$ on the $\alpha$ axis ($\alpha=x,y,z$) and $D_{ij}$ is the coupling constant of the DDI between spins $i$ and $j$. For a linear one-dimensional spin chain $D_{ij}$ is 
   \begin{equation}
    D_{ij}=\frac{\gamma^2 \hbar (1-3\cos^2\theta)}{2a^3|i-j|^3},
   \end{equation}
	where  $\gamma$ is the gyromagnetic ratio, $a$ is the distance between nearest neighbors, and $\theta$ is the angle between the vector connecting spins $i$, $j$ and axis $z$, which is the direction of the strong external magnetic field.
	
	To perform MQ NMR experiments we used quasi-one-dimensional chains of  \BPChem{\^{19}F} nuclei in calcium fluorapatite \BPChem{Ca\_5(PO\_4)\_3F} \cite{Doronin3}. 
Experiments were performed on a Brucker Avance III spectrometer with a static magnetic field $B_0=9.4$ T (the corresponding frequency on \BPChem{\^{19}F} nuclei is 376.6 MHz). 
The main block of the MQ NMR experiment (Fig. \ref{fig:MQNMR}) is repeated many times with the phase  increment of the radio-frequency pulses irradiating the spin system on the preparation period at each repetition. The applied sequence of radio-frequency  pulses is built of basic cycles consisting of eight resonance (for \BPChem{\^{19}F} nuclei) $\pi/2$ pulses (Fig. \ref{fig:MQNMR}). At the preparations period of the MQ NMR experiment, the system is irradiated by several such cycles. At the phase increment of the radio-frequency pulses $\Phi$, the nonsecular two-spin/two-quantum Hamiltonian (\ref{averaged}) has the  form (Fig. \ref{fig:MQNMR})
   \begin{equation}
   \label{twotwo}
    H^{(0)}_\Phi=e^{-2i\Phi}H^{(+2)}+e^{2i\Phi}H^{(-2)}.
    \end{equation}
The phase increments of the radio-frequency pulses serve the purpose of the separation of signals from the MQ NMR coherences of different orders on the free evolution period \cite{Shykind} (Fig. \ref{fig:MQNMR}).

Since the MQ NMR coherences cannot be observed directly they are transformed into the transverse magnetization (single-quantum coherence) in the mixing period. During the mixing period the spin system is irradiated with the same sequence of the radio-frequency pulses as during the preparation period but the pulse phase is shifted by $\pi/2$ ($y$-pulses are applied instead of $x$-pulses). 
Formula (\ref{twotwo}) shows that the sign of the nonsecular two-spin/two-quantum Hamiltonian (\ref{averaged}) changes due to such a phase shift. 
The time reversal conditions \cite{Rhim} are necessary for the co-phasing of contributions of different pairs of spins into the MQ NMR coherences of different orders \cite{Baum}.

The detecting pulse is applied on the detection period for the transmission of the magnetization in the plane perpendicular to the external magnetic field. 
 The Fourier transform with respect to the phase increment makes it possible to obtain MQ NMR spectrum consisting of a set of narrow lines corresponding to different coherence orders \cite{Shykind}.

We emphasize that only MQ NMR coherences of the zeroth and plus/minus second orders emerge in the model of the isolated spin chains in the approximation of the nearest neighbor interactions \cite{Feld2, Feld3,Doronin2, Feld4}.
Our experimental work \cite{Doronin3} demonstrates that more than 97 \% of the observed signal corresponds to the MQ NMR coherences of the zeroth and plus/minus second orders. 
The theoretical intensities of the MQ NMR coherences of the zeroth ($G_0(\tau)$) and plus/minus second ($G_{\pm2}(\tau)$ orders on the preparation period in infinite spin chains are the following \cite{Feld2, Feld3,Doronin2, Feld4}:
\begin{eqnarray}
\label{infinite}
G_0(\tau)=\frac{1}{2}+\frac{1}{2}J_0(4D\tau),\\
G_{\pm2}(\tau)=\frac{1}{4}- \frac{1}{4} J_0(4D\tau),\nonumber
\end{eqnarray}
where $J_0$ is the zero-order Bessel function and $D=D_{i,i+1}=16.4\cdot10^3$~rad/s is the DDI nearest neighbor coupling constant in the spin chain of \BPChem{\^{19}F} nuclei in calcium fluorapatite when the external magnetic field is directed along the spin chain.
\section{A quantum state transfer in one-dimensional spin chains}
The ideal quantum state transfer is possible only for homogeneous spin chains consisting of up to 3 spins \cite{Christandl}.
However, one can engineer the coupling constants \cite{Christandl2} or add external manipulation of the spins at the chain ends \cite{Burgarth} in order to perform the  perfect transfer for chains of arbitrary length.
The quantum state transfer in spin chains in the approximation of the nearest neighbors is performed when spin dynamics is governed by the flip-flop Hamiltonian $H_{ff}$ which can be written as
\begin{equation}
\label{flipflop}
H_{ff}=\sum\limits_i D_{i,i+1} (I^+_i I^-_{i+1}+I^-_i I^+_{i+1}).
\end{equation}

Unfortunately, such a Hamiltonian is not the Hamiltonian in MQ NMR experiments. However, the Hamiltonian (\ref{flipflop}) is related to the two-spin/two-quantum Hamiltonian (\ref{averaged}) by a very simple unitary transformation \cite{Doronin}. Indeed, performing the transformation $U$ of the Hamiltonian (\ref{averaged})
\begin{equation}
\label{unitary}
U=\exp(-i\pi I_{2x})\cdot \exp(-i\pi I_{4x})\dots,
\end{equation}

We apply selective $\pi$ pulses which flip the spins in even positions through $180^\circ$ about the $x$-axis of the rotating  reference frame.
As a result, we obtain
\begin{equation}
U H^{(0)}U^+=\sum\limits_i D_{i,i+1} (I_i^+ I_{i+1}^- +I_i^- I_{i+1}^+)=H_{ff}.
\end{equation}
Although the transformation (\ref{unitary}) cannot be realized experimentally it is very useful for the understanding of peculiarities of the quantum state transfer along spin chains.

The  system's initial state $\rho(0)$, which we consider,  has just spin $l$ polarized \cite{Cappellaro2}
    \begin{equation}
     \label{polarized}
     \rho(0)=\frac{1}{Z} e^{\beta I_{l z}}\otimes E_{N\neq l},
     \end{equation}
		where the parameter $\beta$ is inversely proportional to the temperature, $N$ is the number of  spins in the chain, $Z=2^N\ch{\frac{\beta}{2}}$ is the partition function, $l$ is  a positive  odd number and $E_{N\neq l}$ is the identity operator in the Hilbert space of all spins besides spin $l$.
		In the MQ NMR experiment the density matrix $\rho(t)$ of the system at time   $t$ is
		\begin{equation}
		\rho(t)=e^{-iH^{(0)}t} \rho(0) e^{i H^{(0)}t}.
		\end{equation}
The polarization of spin $N$ at time  $t$ is
\begin{align}
\label{polarization}
   \langle I_{Nz}\rangle &=\Tr\{\rho(t)I_{Nz}\}=
	 \Tr\{U\rho(t)I_{Nz}U^+\}=\\
	& \qquad\Tr\{U\rho(t)U^+UI_{Nz}U^+\}= \Tr\{e^{-iH_{ff}t}\rho(0)e^{iH_{ff}t }I_{Nz}\}.\nonumber
	\end{align}
In Eq.~(\ref{polarization}) we suppose that $N$ is also odd. Thus, we have shown that the quantum state transfer in the MQ NMR experiment can be described by the flip-flop Hamiltonian of Eq.~(\ref{flipflop}).

The spin dynamics of the considered system is solved exactly in terms of fermionic operators and their Fourier transform \cite{Doronin2,Feld2}. If the initial polarization was on spin $l$ (see Eq. (\ref{polarized})) than the polarization of spin $m$ at time $t$ is \cite{Feld5}
\begin{equation}
\label{inpo}
\frac{\left \langle I_{mz}\right\rangle (t)}{\left\langle  I_{lz} \right\rangle (t)}=\frac{4}{(N+1)^2}\left |\sum_k \exp(-i\epsilon_k t)\sin(kl)\cdot\sin(km)\right|^2,
\end{equation}
where  the fermion spectrum $\epsilon_k$ is
\begin{equation}
\epsilon_k=D\cos(k)+\omega_0,
\end{equation}
where $D=D_{i,i+1}$, $\omega_0$ is the Larmor frequency and
\begin{equation}
k=\frac{\pi n}{N+1}, \qquad n=1,2,\dots N.
\end{equation}
Eq. (\ref{inpo}) was obtained \cite{Feld5} in the high temperature approximation \cite{Abragam}.
However, it was  shown \cite{Feld6} that this relationship holds at arbitrary temperatures.

Numerical calculations show that almost all polarization (more than 90 \%) can be transmitted to the given chain site for short chains. 
For long chains the efficiency of transmission gets lower.

The developed approach to the quantum state transfer relies on the approximation of the nearest neighbor interactions.
It is known \cite{Cappellaro} that the DDI of remote spins are not important for the quantum state transfer in short chains.

\section{Relaxation of MQ NMR coherences as a model of decoherence processes}
The dipolar relaxation of MQ NMR coherences can be studied on the evolution period of the MQ NMR experiment.
Relaxation of MQ NMR coherences  in one-dimensional systems has been studied earlier \cite{Doronin,Kaur} using the second moments of the line shapes of the MQ coherences of the zeroth and second orders.
We suggested \cite{Bochkin} to perform an analogous investigation using the ZZ-model, where only the ZZ-part of the DDI is taken into account.

The exact solution \cite{Feld2,Feld3,Doronin2,Feld4} for MQ NMR dynamics in one-dimensional systems allows us to obtain the density matrix $\sigma(\tau)$ on the preparation period of the MQ NMR experiment \cite{Baum} in the approximation of the nearest neighbor interactions \cite{Abragam,Goldman} and to write it as follows
\begin{equation}
\label{preparation}
\sigma(\tau)=\sigma_0(\tau)+\sigma_2(\tau)+\sigma_{-2}(\tau),
\end{equation}
where $\sigma_i(\tau)$ ($i=0,2,-2$) describes the MQ NMR coherence of order $i$.
If the number of the spins $N\gg 1$ the contributions $\sigma_0(\tau)$, $\sigma_2(\tau)$, $\sigma_{-2} (\tau)$ are \cite{Feld2,Feld3}
\begin{eqnarray}
\label{sigma0}
\sigma_0 (\tau)&=&\frac{1}{2}\sum_k\cos[2 D \tau \sin(k)](1-a_k^+a_k),\\
\sigma_2(\tau)&=&-\frac{1}{2}\sum_k \sin[2 D \tau \sin(k)]a_k a_{-k},\\
\sigma_{-2}(\tau)&=&\frac{1}{2}\sum_k \sin[2 D \tau \sin(k)]a_k^+ a_{-k}^+,
\end{eqnarray}
where $k=\frac{2\pi n}{N}$, ($n=-N/2,\,-N/2+1,\,\dots N/2-1$) and $a_k^+$, $a_k$ are the fermion operators \cite{Feld2,Feld3}.

We study the dipolar relaxation of the MQ NMR coherences \cite{Bochkin} on the evolution period of the MQ NMR experiments \cite{Baum}. 
This period begins immediately after the preparation period \cite{Baum} and the density matrix (\ref{preparation}) can be used as the initial state for the relaxation.
The relaxation of MQ NMR coherences is caused by the secular (with respect to the external magnetic field directed along the $z$-axis) DDI
\begin{equation}
\label{ham}
H_{dz}=\sum_{i<j} D_{ij} (3 I_{iz} I_{jz} - \vec{I_i} \vec {I_j})=\sum_{i<j}D_{ij}(2I_{iz} I_{jz} - I_{ix} I_{jx}-I_{iy}I_{iy})
\end{equation}
An investigation of the relaxation process with the Hamiltonian (\ref{ham}) is too complicated. 
However, the problem is substantially simplified \cite{Bochkin}, if we restrict ourselves to the ZZ-part of  $H_{dz}$ only and consider the Hamiltonian
\begin{equation}
H_{ZZ}=2 \sum_{i<j} D_{ij} I_{iz} I_{jz}=\sum_{i\neq j} D_{ij} I_{iz} I_{jz}.
\end{equation}
The problem with the Hamiltonian $H_{ZZ}$ was called \cite{Bochkin}  the ZZ-model.

This approximation is widely used \cite{Abragam} in spin physics (for example, see \cite{Feld7}).
The intensity $F_0(\tau,t)$ of the MQ NMR coherence of the zeroth order in the course of the evolution period at the time moment $t$ is

\begin{equation}
\label{evolution}
F_0(\tau,t)=\frac
               {\Tr\{e^{-iH_{ZZ}t}\sigma_0(\tau) e^{iH_{ZZ}t} \sigma_0(\tau)\}}
                                                                                      {\Tr(I_z^2)}=
					\frac{\Tr\{e^{-iH_{ZZ}t}\sigma_0(\tau) e^{iH_{ZZ}t} \sigma_0(\tau)\}}{N\cdot 2^{N-2}}.																																						
\end{equation}
In Eq. (\ref{evolution}) $I_z=\sum_{i=1}^N I_{iz}$ and $F_0(\tau,0)=G_0(\tau)$ (see Eq. (\ref{infinite})).

Analogously, the intensities of the MQ NMR coherences of the plus/minus second order on the evolution period are given by

\begin{equation}
F_{\pm 2} (\tau,t)=\frac{\Tr\{e^{-iH_{ZZ}t}\sigma_2(\tau) e^{iH_{ZZ} t} \sigma_{-2}(\tau)\}}{\Tr(I_z^2)}.
\end{equation}

\begin{figure}
\begin{center}
\begin{tabular}{c}
\includegraphics[height=5.5cm]{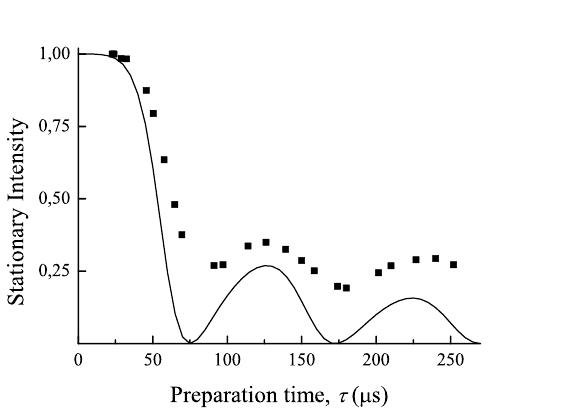}
\end{tabular}
\end{center}
\caption 
{ \label{fig: intensity}
The stationary intensity of the MQ NMR coherence of the zeroth order on the preparation period. The solid line is the theoretical plot of Eq. (\ref{stationary}).} 
\end{figure}

Our calculations demonstrate \cite{Bochkin} that the MQ NMR coherence of the zeroth order is not subject to the dipolar relaxation in the ZZ-model. However, the experimental investigation \cite{Bochkin} performed using quasi-one-dimensional chains of  \BPChem{\^{19}F} nuclei in calcium fluorapatite \BPChem{Ca\_5(PO\_4)\_3F} shows the dipolar  relaxation of the MQ NMR coherence of the zeroth order due to the flip-flop part of the Hamiltonian (\ref{ham}). The experimental data \cite{Bochkin} also demonstrate that relaxation does not lead to the full disappearance of the MQ NMR of the zeroth order. 
Relaxation ends with the stationary intensity of that MQ coherence.
The point is that the density matrix of Eq. (\ref{sigma0}) contains a part which is proportional to the operator $I_z$, commuting with the DDI Hamiltonian (\ref{ham}). As a result, one can obtain \cite{Bochkin} the stationary intensity of the MQ NMR coherence of the zeroth order $F_o^{st}$ which is
\begin{equation}
\label{stationary}
F_0^{st}=\frac{J_0^2(2D\tau)}{G_0(\tau)}=\frac{2J_0^2(2D\tau)}{1+J_0(4D\tau)}.
\end{equation}
The experimental dependence of the stationary intensity of the MQ NMR coherence of the zeroth order during  the preparation period is presented in Fig. \ref{fig: intensity}. The experimental data are in a close agreement with Eq. (\ref{evolution}).

\begin{figure}
\begin{center}
\begin{tabular}{c}
\includegraphics[height=5.5cm]{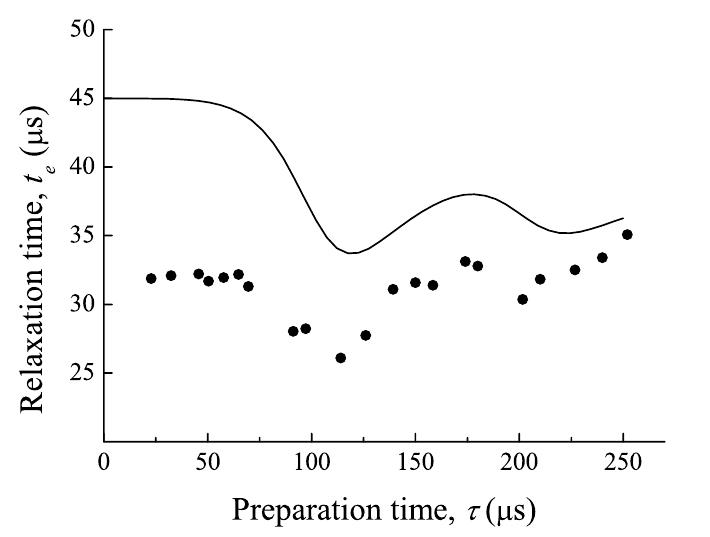}
\end{tabular}
\end{center}
\caption 
{ \label{fig:time}
The experimental times of the dipolar relaxation of the MQ NMR coherence of the second order versus the duration of the preparation period $\tau$. The solid line shows the theoretical times $t_e$ of the dipolar relaxation obtained as $t_e=\sqrt{\frac{2}{M_2(\tau)}}$, where  the second moment $M_2$ is determined by Eq. (\ref{second}).}
\end{figure}

Calculations of the relaxation of the MQ NMR coherences of the plus/minus second order with Eq. (\ref{evolution}) yield \cite{Bochkin}
\begin{equation}
F_{\pm2}(\tau,t)=\frac{1}{8N}\sum _{m, m'}\prod_{n\neq(m,m')}\cos [(D_{nm}+D_{nm'})t] \cdot [1-(-1)^{m-m'}]^2J^2_{m-m'}(2D\tau).
\end{equation}
We have shown \cite{Bochkin} that the decay of the MQ NMR coherences of the second order conforms reasonably well to a Gaussian function 
$S(t)=\exp\{-\frac{M_2(\tau) t^2}{2}\}$, where the second moment $M_2(\tau)$ of the line shape of this coherence is 
\begin{equation}
\label{second}
M_2(\tau)=-\left.\frac{1}{G_2(\tau)} \frac{d^2 F_{\pm2}(\tau,t)}{dt^2}\right|_{t=0}
\end{equation}
In Fig. \ref{fig:time} the dependencies of the experimental and theoretical (at $N=150$) dipolar relaxation times on the duration of the preparation period are shown. 
One can see that the theoretical predictions satisfactorily describe the experimental data.

\section{Conclusions}
We considered the methods of MQ NMR spectroscopy of one-dimensional spin systems and their application for solving problems of quantum informatics.
The quantum state transfer in linear spin chains can be realized with high fidelity using MQ NMR dynamics.
Decoherence of many-qubit quantum state created on the preparation period of the MQ NMR experiment \cite{Baum} can be studied on the evolution period of that experiment. 
The dipolar relaxation of the MQ NMR coherences is considered as a model of the decoherence process.
The theory of the dipolar relaxation of the MQ NMR coherences is developed.
The performed MQ NMR experiments on the quasi-one-dimensional system of  \BPChem{\^{19}F} nuclei in calcium fluorapatite \BPChem{Ca\_5(PO\_4)\_3F} are in a good agreement with the theoretical predictions.

\acknowledgments
The work is supported by the Russian Foundation for Basic Research (Grants 16-03-00056 and 16-33-00867) and the Program of RAS "Element base of quantum computers" (Grant No 0089-2015-0220).

\end{document}